# Pyroelectric response of ferroelectric nanoparticles: size effect and electric energy harvesting

A.N. Morozovska [a*], E.A. Eliseev [b], G.S. Svechnikov [a], and S.V. Kalinin[c†]

[a]V. Lashkarev Institute of Semiconductor Physics, National Academy of Sciences of Ukraine,

41, pr. Nauki, 03028 Kiev, Ukraine

[b]Institute of Problems of Materials Science, National Academy of Sciences of Ukraine,

3, Krjijanovskogo, 03142 Kiev, Ukraine

[c]Oak Ridge National Laboratory, Oak Ridge, TN 37831

**Abstract**

The size effect on pyroelectric response of ferroelectric nanowires and nanotubes is analyzed. The pyroelectric coefficient ***strongly increases*** with the wire radius decrease and diverges at critical radius $R_{cr}$ corresponding to the size-driven transition into paraelectric phase. Size-driven enhancement of pyroelectric coupling leads to the giant pyroelectric current and voltage generation by the polarized ferroelectric nanoparticles in response to the temperature fluctuation. The maximum efficiency of the pyroelectric energy harvesting and bolometric detection is derived, and is shown to approach the Carnot limit for low temperatures.

## I. Introduction

The unique features of nanosized piezo-, pyro- and ferroelectrics enable a broad spectrum of thermo-electrical, electro-mechanical, electronic and dielectric properties for sensors and actuators, compact electronics, pyrosensors and thermal imaging [1, 2]. Piezoelectric nanowires have been studied as potential strain – based energy harvesting devices, in particular as a direct current generators [3, 4, 5].

---

[*] morozo@i.com.ua
[†] sergei2@ornl.gov

In these, piezoelectric nanowires array is aligned normally to substrate and sandwiched between the bottom substrate electrode and the top ratchet-like electrode. The acoustic excitation of piezoelectric nanowires leads to their bending and results in charge generation. The output *d.c.* current value is in the range of several pA to nA, which is enough for different nanoscale devices power supply. However, the presence of moving parts may result in rapid degradation of the structure.

An alternative approach for ferroelectric-based energy harvesting is based on the pyroelectric properties. Early attempts [6, 7] showed the efficiency of pyroelectric energy conversion is about ten percents for polymeric and ceramic bulk ferroelectrics. Recently Mischenko et al. [8, 9] demonstrated the giant electrocaloric effect, i.e. the temperature change in response to electric field applied under adiabatic conditions, in $PbZr_{0.95}Ti_{0.05}O_3$ films [8] and relaxor ferroelectric $0.9\ PbMg_{1/3}Nb_{2/3}O_3 - 0.1\ PbTiO_3$ [9] near the ferroelectric Curie temperatures (correspondingly 222°C and 60°C). Mischenko et al. pointed out that the direct electrocaloric effect and pyroelectric effects are strongly enhanced in the vicinity of phase transitions, potentially enabling efficient Peltier-type devices and efficient energy harvesting approaches.

The rapid progress in synthesis of ferroelectrics nanoparticles, in particular vertical arrays of free-standing tubes [10], wires [11] and rods in porous template [12, 13], demonstrated their enhanced polar properties and unusual domain structure. Then possibility to control the temperature of the phase transitions in ferroelectric nanoparticles due to the size-driven phase transition has been studied theoretically [14, 15, 16, 17, 18].

The size effect can be used to tune the phase transition temperatures in ferroelectric nanostructures, thus enabling the systems with tunable giant pyroelectric response. However, unlike the well-studied theoretically ferroelectric and dielectric properties, dynamic behavior of pyroelectric response of ferroelectric nanostructures has not been considered, and a small number of existing treatments have been limited to pyroelectric coefficient calculations in thin epitaxial films [19, 20] and rods [21]. Here, we analyze in details the dynamic pyroelectric response of polarized ferroelectric nanotubes and wires within Landau-Ginzburg-Devonshire phenomenological theory, and derive the power spectrum and efficiency of idealized devices (arrays of ferroelectric **nanowires** or **nanotubes** fixed in a flat capacitor). We demonstrate that the devices are prominent candidates for the ***harvesting of electric energy*** from different heat sources.

## II. Basic equations

Bound charge excess $Q_{pr}$ (typically called "pyroelectric" charge) appears at the polar faces of ferroelectric in response to the time-dependent thermal flux via the temperature variation $d\delta T/dt$ (see Fig. 1a). Hereinafter the variation $\delta T(t)$ is regarded spatially quasi-homogeneous across the nanoparticle and small in comparison with ambient temperature $T_0$, namely: $T(\mathbf{r},t) = T_0 + \delta T(\mathbf{r},t)$ and $T_0 \gg |\delta T(\mathbf{r},t)|$. The temperature fluctuations $d\dot T/dt$ have a known frequency spectrum $\delta \dot T(\omega) = \int_0^\infty dt (dT/dt) e^{i\omega t}$.

The distinctive feature of the nanosized systems pyroelectric response is the spatial inhomogeneity of the pyroelectric charge originated from the spatial inhomogeneity of their spontaneous polarization distribution $P_3$ related with the surface influence on elementary dipoles correlation (see Fig. 1a).

Pyroelectric coefficient is given by expression [1]:

$$\Pi_3(T_0) = \left( \frac{\partial P_3}{\partial T} + \left( \frac{\partial P_3}{\partial u_{ij}} \right) \cdot \frac{\partial u_{ij}}{\partial T} \right)_{T=T_0}. \qquad (1)$$

The first term $\partial P_3/\partial T$ originated from the primary pyroelectric effect related to spontaneous polarization changes, the second term $(\partial P_3/\partial u_{ij})(\partial u_{ij}/\partial T)$ originated from the secondary pyroelectric effect related with the possible temperature dependence of mechanical strains $u_{ij}$ via thermal expansion. In Eq.(1) we neglected the ternary pyroelectric effect originated from inhomogeneous temperature distribution that leads to polarization variation as $\delta P_i = f_{ijkl} a_{kl}^T (\partial T/\partial x_j)|_{T=T_0}$ ($f_{ijkl}$ is the flexoelectric effect tensor, thermal expansion coefficients are denoted as $a_{ij}^T$). Under low enough $\partial T/\partial x_j$, the ternary effect is negligibly small in pyroelectrics and ferroelectrics in comparison with the primary and even secondary effects. Below we calculate pyroelectric response of ferroelectric nanoparticles within Landau-Ginzburg-Devonshire phenomenological theory.

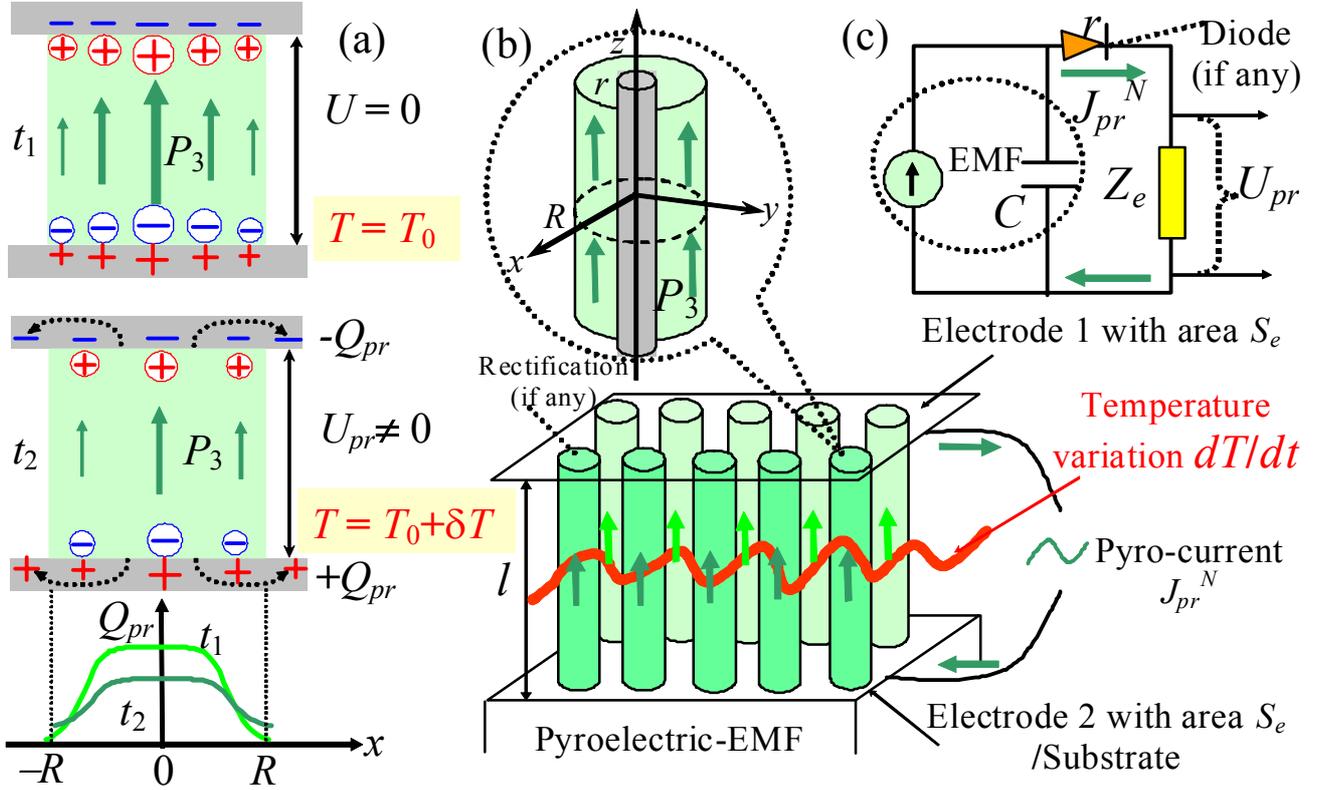

**FIG. 1**. (Color online) (a) Generation of the pyroelectric charge $Q_{pr}$ in the electrodes around a ferroelectric nanorod with inhomogeneous spontaneous polarization $P_3$ under its temperature variation on $\delta T(t)$. The lower plot schematically shows the charge $Q_{pr}$ profile across the rod at time moments $t_1$ and $t_2$. (b) Pyroelectric electromotive force (EMF) and current generation by ferroelectric nanostructures: a single polarized rod (with or without rigid core) and nanowires vertically-aligned array in contact with the electrode plates. (c) Equivalent circuit of operating pyroelectric electromotive force: $C$ is the effective capacitance of the capacitor, $r$ is the diode (if any appeared due to the possible rectification effect at the ferroelectric-semiconductor-electrode interface), $Z_e$ is the external load impedance, at that $J_{pr}^n(\omega) \approx U_{pr}/Z_e - i\omega C U_{pr}$ for small internal resistance of current source.

Correct phenomenological description of nanosized system requires the consideration of appropriate surface energy. Including the surface energy term $F_S$, Landau-Ginzburg-Devonshire free energy $F$ depends on the chosen order parameter – spontaneous polarization component $P_3$ and mechanical strains $u_{ij}$ as [15]:

$$F = \int_S d^2r \frac{\alpha_S}{2} P_3^2 + \int_V d^3r \left( \frac{\alpha}{2} P_3^2 + \frac{\beta}{4} P_3^4 + \frac{\gamma}{6} P_3^6 + \frac{g}{2} (\nabla P_3)^2 - P_3 \left( E_e + \frac{E_3^d}{2} \right) - q_{ij33} u_{ij} P_3^2 + \frac{c_{ijkl}}{2} u_{ij} u_{kl} \right) \quad (2)$$

Typically the surface energy coefficient $\alpha_S$ is regarded positive, isotropic and weakly temperature dependent, thus the terms $\sim P_3^4$ can be neglected in the surface energy expansion. Integration in the first and the second terms of Eq. (2) is performed over the system surface $S$ and volume $V$ correspondingly. Expansion coefficient $\beta > 0$ for the second order phase transitions, $\gamma > 0$ and the gradient coefficient $g > 0$. Coefficient $\alpha(T) = \alpha_T (T - T_C)$, $T$ is temperature; $T_C$ is Curie temperature of bulk material. The stiffness tensor $c_{ijkl}$ is positively defined, $q_{ijkl}$ stands for the electrostriction stress tensor. $E_e$ is the external electric field. Considering high aspect ratio cylindrical nanoparticles (high aspect ratio nanoellipsoids, nanotubes or nanowires with length $l$ much higher than radius $R$) with spontaneous polarization directed along the cylinder axes z we will neglect the effects of depolarization field $E_d$ in Eq.(1).

Minimization of the free energy on polarization and strain components gives the equations of state. These equations are supplemented with Maxwell equations for electrostatic electric field and compatibility conditions for strain and equilibrium conditions $\partial \sigma_{ij} / \partial x_i = 0$ for stress components.

The intrinsic surface stress $\mu_{\alpha\beta}^S$ exists under the curved surface of solid body and determines the excess pressure on the surface [14-17]. The surface stress tensor $\mu_{\alpha\beta}^S$ is defined as the derivative of the surface energy on the deformation tensor. Intrinsic mechanical stress under curved surface is determined by the tensor of intrinsic surface stress as $n_k \sigma_{kj} \big|_S = -n_j \mu_{\alpha\alpha}^S / R_\alpha$, where $R_\alpha$ are the main curvatures of surface free of facets and edges in continuum media approximation, $n_k$ are the components of the external normal.

The strain field inside cylindrical ferroelectric nanoparticles is rather complicated because of the spatially distributed polarization. Using the Saint-Venant principle one could get the quantitatively correct physical picture (except the immediate vicinity of the faces $z = 0$ and $z = l$) and derived appropriate analytical expressions for the strain field $u_{ij}$ in nanoparticles (see Appendix).

Variation of the free energy functional (1) leads to the Euler-Lagrange equations:

$$\begin{cases} \alpha_R P_3 + \beta_R P_3^3 + \gamma P_3^5 - g\Delta P_3 = E_0, \\ \left. \left( P_3 + \lambda \dfrac{\partial P_3}{\partial \mathbf{n}} \right) \right|_{\mathbf{r} \in S} = 0 \end{cases} \qquad (3)$$

Where $\Delta$ is Laplace operator. Note that the polarization relaxation time is extremely small (about $10^{-10}$ s) in comparison with the temperature rates $dT/dt \sim 0.01\text{-}1$ K/s. Hence, we omit the time derivatives in the Euler-Lagrange equation (3). Extrapolation length $\lambda = g/\alpha_S$ is positive, $\mathbf{n}$ is the outer normal to the surface $S$.

For tetragonal ferroelectric and cubic elastic symmetry groups coefficients $\alpha$ and $\beta$ are renormalized by thermal expansion, surface tension and strains as:

$$\alpha_R = \alpha_T (T - T_C) + q_{ij33} a_{ij}^T (T - T_0) + \frac{4 Q_{12} \mu^S}{R} - \frac{u_c r^2}{\alpha_T R^2}\left( q_{11} + \left(1 - 2\frac{c_{12}}{c_{11}}\right) q_{12} \right), \qquad (4a)$$

$$\beta_R \approx \beta - 2\frac{q_{12}^2}{c_{11}} \frac{r^2}{R^2} - 2\frac{(c_{11} + c_{12}) q_{11}^2 - 4 c_{12} q_{11} q_{12} + 2 c_{11} q_{12}^2}{(c_{11} - c_{12})(c_{11} + 2 c_{12})}\left(1 - \frac{r^2}{R^2}\right). \qquad (4b)$$

The tube outer radius is $R$, the inner radius is $r$ (see Fig. 1b); $u_c$ is the strain at the interface $\rho = r$ (if any). For the practically important case of the ferroelectric tube deposited on a rigid dielectric core, the tube and core lattices mismatch or the difference of their thermal expansion coefficients determines $u_c$ value (allowing for the possible strain relaxation for thick tubes). Parameter $Q_{12} = \dfrac{c_{11} q_{12} - c_{12} q_{11}}{(c_{11} - c_{12})(c_{11} + 2 c_{12})}$ stands for the stress electrostriction coefficient. The second terms in Eqs.(4) originated from thermal expansion $\sim a_{ij}^T$, the third terms originated from intrinsic surface stress $\sim \mu^S$, the third terms are the strains induced by core $\sim u_c$, the last terms are the spontaneous strains created by inhomogeneous polarization.

Using direct variational method [14-15], the approximation for the averaged spontaneous polarization $\overline{P}_3$ and dielectric susceptibility $\overline{\chi}_{33}$ were derived as:

$$\overline{P}_3(R, r, T_0) = \sqrt{\frac{2\alpha_T (T_{cr}(R, r) - T_0)}{\beta_R + \sqrt{\beta_R^2 + 4\gamma \alpha_T (T_{cr}(R, r) - T_0)}}}, \qquad (5a)$$

$$\overline{\chi}_{33}(R, r, T_0) = \frac{1}{2\alpha_T (T_{cr}(R, r) - T_0) + 3\beta \overline{P}_3^2 + 5\gamma \overline{P}_3^4}. \qquad (5b)$$

Hereinafter a dash over the letter stands for the averaging over the nanoparticle volume. Corresponding temperature of the size-driven transition to the paraelectric phase acquires the form:

$$T_{cr}(R,r) = T_C - \frac{4Q_{12}\mu^S}{\alpha_T R} + \frac{u_c r^2}{\alpha_T R^2}\left(q_{11} + \left(1 - 2\frac{c_{12}}{c_{11}}\right)q_{12}\right) - \frac{2g}{\alpha_T}\left(\lambda(R-r) + \frac{2R(R-r)^2}{\pi^2 r + k_0^2(R-r)}\right)^{-1}. \quad (6)$$

Constant $k_0 = 2.408$ is the minimal root of Bessel function $J_0(k)$. By changing the wire radius one can tune the transition temperature in the wide range. The first term in Eqs.(6) is the bulk transition temperature, the second one is the contribution of intrinsic surface stress $\sim\mu^S$, the third term is the effect of mismatch strain $u_c$, the last term $\sim g$ originated from correlation effects.

The size-dependent pyroelectric coefficient is calculated from Eq.(1) as:

$$\overline{\Pi}_3(T_0) \approx -\sqrt{\frac{\alpha_T\left(\beta_R + \sqrt{\beta_R^2 + 4\gamma\alpha_T(T_{cr} - T_0)}\right)}{8(T_{cr} - T_0)\left(\beta_R^2 + 4\gamma\alpha_T(T_{cr} - T_0)\right)}}\left(1 + q_{ij33}\frac{a_{ij}^T}{\alpha_T}\right). \quad (7)$$

For the sake of simplicity we considered the case of the nanowire without core (i.e. $r = 0$). The pyroelectric coefficient increases with the wire radius $R$ decrease and diverges at critical radius $R_{cr}$ corresponding to the size-driven transition into paraelectric phase and then drops to zero in paraelectric phase (see Figs. 2a,b). The value of $R_{cr}$ is determined from the condition of susceptibility divergence $\overline{\chi}_{33}(R_{cr}, T) \to \infty$. Thus it is possible to tune the pyroelectric coefficient value by varying the nanowire radius $R$ and ambient media characteristics responsible for surface tension coefficients $\mu^S$ and surface energy coefficient $\alpha_S$ (since $\lambda^{-1} \sim \alpha_S$).

The physical characteristics of the nanowire array for should be averaged over their radii $R$ with corresponding distribution function $f(R)$ as $\langle F \rangle = \int_{R_{min}}^{R_{max}} F(R)f(R)dR$. The averaging leads to the noticeable smearing of the size dependences of pyroelectric coefficient and susceptibility, at that the divergences at critical radius transform into maxima and to the appearance of the dispersion of pyroelectric coefficient and susceptibility maxima position corresponding to different halfwidth $\delta R$ of the sizes distribution functions (see Fig. 2c). The smearing and dispersion increase with relative halfwidth $\delta R/\langle R \rangle$ increase (compare the curves 0, 1, 3 and 4).

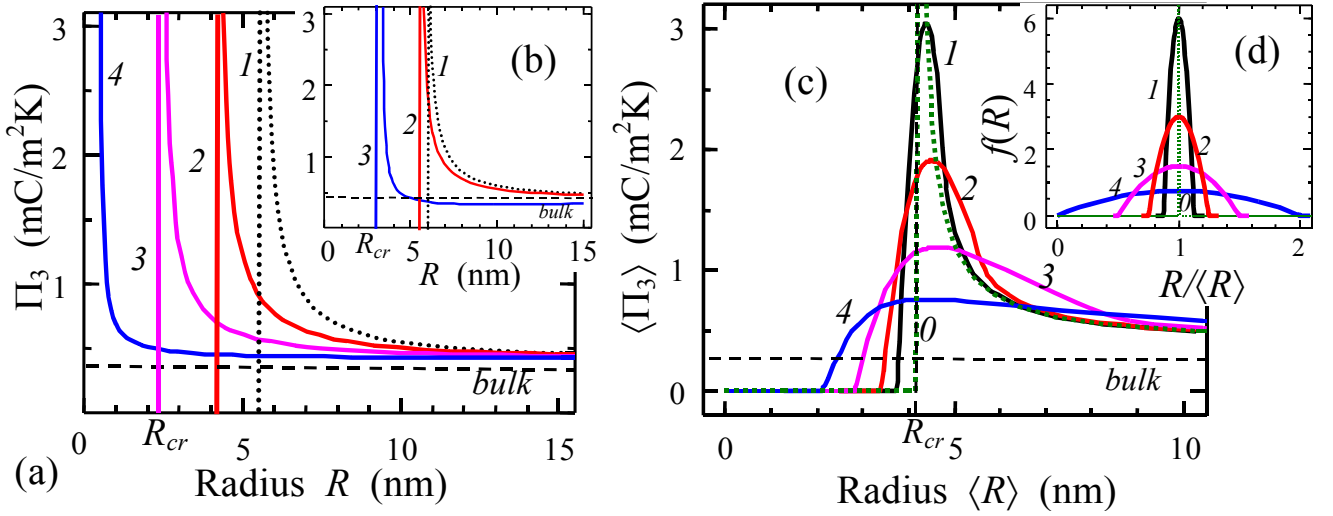

**FIG. 2**. (Color online) Pyroelectric coefficient $\Pi_3$ vs. nanowire radius $R$ for different surface tension coefficients $\mu^S = 0, 1, 10$ N/m and fixed length $\lambda = 0$ (curves 1-3 in plot (b)); different length $\lambda = 0, 1, 3, 10$ nm and fixed surface tension coefficients $\mu^S = 1$ N/m (curves 1-4 in plot (a)). (c) Averaged pyroelectric coefficient $\langle \overline{\Pi}_3 \rangle$ vs. the average radius $\langle R \rangle$ for different relative halfwidth $\delta R / \langle R \rangle = 0.01$, 0.25, 0.5, 1, 2 (curves 0-4) of the size distribution function $f(R)$ shown in plot (d). Surface tension coefficient $\mu^S = 1$ N/m, length $\lambda = 1$ nm. Horizontal lines indicate $\Pi_3$ of bulk material PbZr$_{0.4}$Ti$_{0.6}$O$_3$. Material parameters of PbZr$_{0.4}$Ti$_{0.6}$O$_3$: $\alpha_T = 4.25 \cdot 10^5$ m/(F K), $T_C = 691$ K, $\beta_R = 1.44 \cdot 10^8$ m$^5$/(C$^2$F), $\gamma = 1.12 \cdot 10^9$ m$^9$/(C$^4$F), $Q_{12} = -0.0295$ m$^4$/C$^2$, room temperature $T = 300°$K, gradient term coefficient $g = 10^{-9}$ m$^3$/F.

Here, we consider the maximal efficiency of pyroelectric energy harvesting device formed by ferroelectric nanowires arranged in vertical array. The pyroelectric current $\overline{J}_{pr}(t) = \pi R^2 \dfrac{d\overline{P}_3}{dt} \equiv \pi R^2 \overline{\Pi}_3 \left( \dfrac{dT}{dt} \right)$ generated by a single ferroelectric nanowire in response to the temperature variation $dT/dt$ has the following power spectrum:

$$\tilde{J}_{pr}(\omega, R) = \delta \dot{T}(\omega) \cdot \overline{\Pi}_3(R, T_0) \cdot \pi R^2. \tag{8}$$

The total current $J^n_{pr}(t)$ and voltage $U_{pr}(t)$ the produced on external loading $Z_e$ by pyroelectric capacitor $C$ filled with the array of $N$ almost identical nanowires vertically-aligned with respect to the electrodes have the following power spectrum:

$$\tilde{J}^n_{pr}(\omega) = \frac{\delta \dot{T}(\omega) \cdot nS_e \cdot \overline{\Pi}_3(R, T_0)}{1 - i\omega C(R, l, n)(Z_e + r)}, \qquad \tilde{U}_{pr}(\omega) = \tilde{J}^n_{pr}(\omega) \cdot Z_e. \qquad (9)$$

The fraction of nanowires in capacitor is defined as $n = N\pi R^2/S_e$, $S_e$ is the electrodes area (see Fig. 1b,c). The effective capacity of the system was estimated as: $C(R,l,n) \approx \varepsilon_0 \varepsilon^{eff}_{33}(R,n) S_e / l$ ($\varepsilon_0$ is universal dielectric constant). For considered geometry effective dielectric permittivity $\varepsilon^{eff}_{33}(R,n) = \varepsilon_e(1-n) + n(1 + \varepsilon_0^{-1} \cdot \overline{\chi}_{33}(R))$ coincides in both self-consistent Bruggeman and Maxwell-Garnett approximations, $\varepsilon_e$ is the ambient dielectric permittivity, dielectric susceptibility of a single nanorod $\overline{\chi}_{33}$ is given by Eq.(5b). The external load has complex impedance $Z_e = r_e - i(\omega L_e - 1/\omega C_e)$.

Let us consider typical exponentially vanishing fluctuation of the rod temperature [22]:

$$\frac{d}{dt}\delta T = \begin{cases} -(\delta T_0/\tau)\exp(-t/\tau), & t > 0, \\ 0, & t \leq 0. \end{cases} \rightarrow \delta \dot{T}(\omega) = -\frac{\delta T_0}{1 - i\omega\tau}. \qquad (10)$$

For the case of active load resistance $Z_e \equiv r_e$ one obtains the time dependences from Eqs.(9) and (10) as

$$J^n_{pr}(t) = \delta T_0 \cdot nS_e \cdot \overline{\Pi}_3(R, T_0) \left( \frac{\exp(-t/(C(r_e+r))) - \exp(-t/\tau)}{C(r_e+r) - \tau} \right)$$ and $U_{pr}(t) = r_e J^n_{pr}(t)$. The current $J^n_{pr}(t)$ and voltage $U_{pr}(t)$ produced by pyroelectric nanowires in response to the temperature fluctuation (10) are shown in Figs. 3 for different radius $R$ and fraction $n$ respectively. It is clear from the Fig. 3a that both pyroelectric voltage and corresponding current $J_{pr} \approx U_{pr}/r_e$ increase with nanowire radius decrease up to the critical value $R_{cr} \approx 5.8$ nm for chosen material parameters and high external resistance $r_e$. Fig. 3b demonstrates the increase of pyroelectric response with increase of nanowire fraction $n$. Additional calculations show that the increase of voltage $U_{pr}$ with radius $R$ decrease is monotonic at constant $n$. Typical form of pyroelectric current impulse in the case of small external resistance $r_e$ is shown in Fig. 3c for different nanowire radius $R$ and fixed fraction $n$. As anticipated the values of $U_{pr}$ and $J_{pr}$ increase linearly with the temperature rate $dT/dt$ increase.

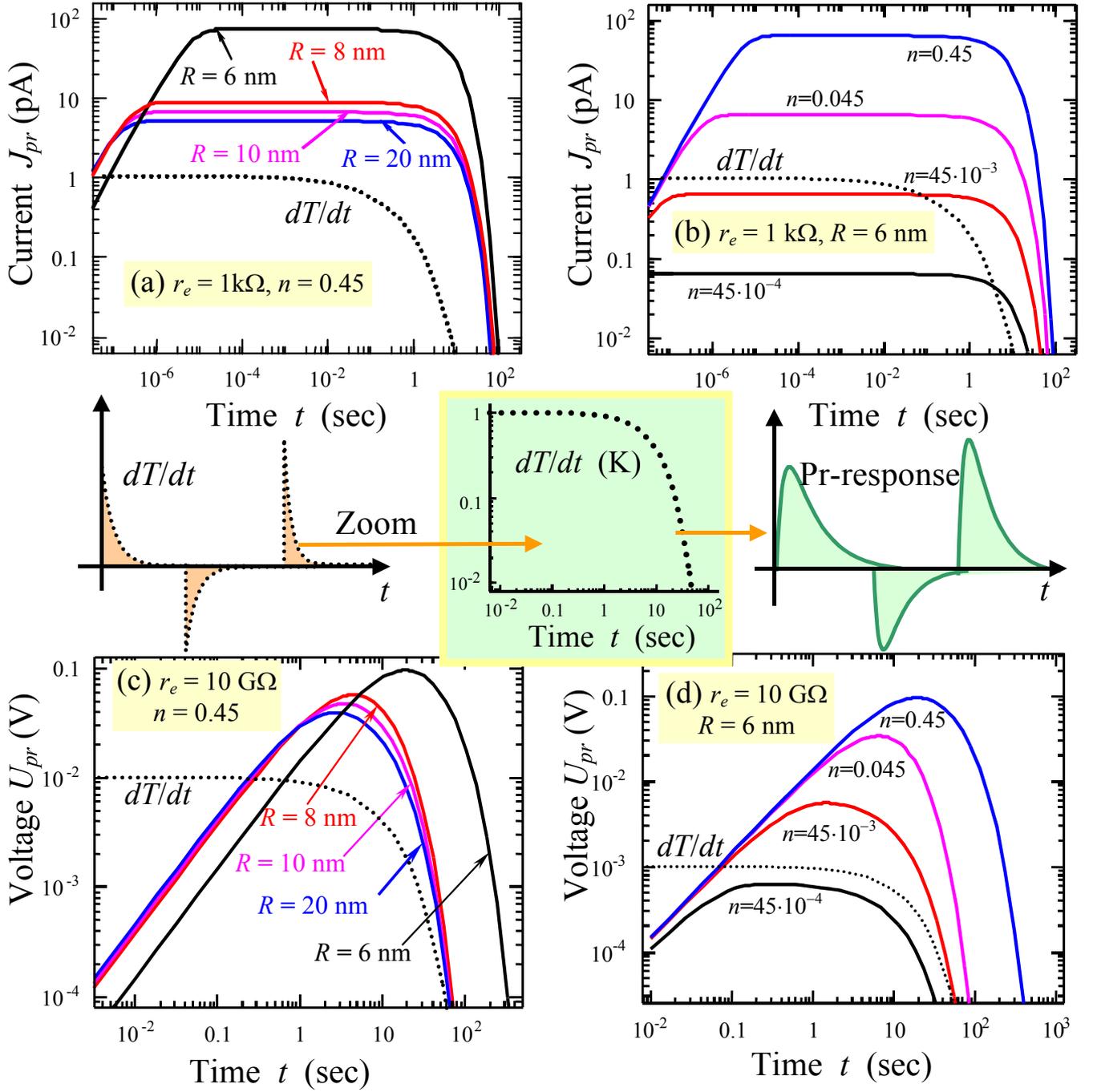

**FIG. 3.** (Color online) Pyroelectric current $J_{pr}^n(t)$ (a,b) and voltage $U_{pr}(t)$ (c,d) vs. time $t$ for different nanowire radius $R = 6, 8, 10, 20$ nm and fixed fraction of nanowires $n = 0.45$ (plots a,c); different fraction of nanowires $n = 45 \cdot 10^{-4}, 4.5 \cdot 10^{-3}, 4.5 \cdot 10^{-2}, 0.45$ and fixed radius $R = 6$ nm (plots b,d), small load resistance $r_e \leq 1$ k$\Omega$ (a,b) and high resistance $r_e = 10$ G$\Omega$ (c,d). Central inset shows the temperature variation $dT/dt$ and its conversion into pyroelectric response. Wires length $l = 1$ μm, electrode area $S_e = 0.25$ mm$^2$, temperature variation amplitude $\delta T_0 = 1$ K, relaxation time $\tau = 10$ s, $\lambda = 0$, $\mu^S = 0$, $\varepsilon_e = 1$. Other parameters are the same as in Fig. 2.

The efficiency η of the power converter is defined as the ratio of electrical work given by the system to the absorbed heat energy. Adopting the calculations made in Ref. [23] for a bulk material to the case of nanoparticle, we obtained the estimation for the actual temperature range $T_0 < T < T_{cr}$:

$$\eta(R,r,T) = \frac{\alpha_T (T_{cr} - T) \overline{P}_3^2}{\alpha_T T_{cr} \overline{P}_3^2 + C_P (T_{cr} - T)} \approx \frac{T_{cr} - T}{T_{cr} + 0.5 \cdot C_P(T) \alpha_T^{-2} \left( \beta_R + \sqrt{\beta_R^2 + 4\gamma \alpha_T (T_{cr} - T)} \right)}. \quad (11)$$

One should take into account the temperature dependence of lattice specific heat, e.g. the Debye law $C_P(T) = 3C_B \left( \frac{T}{\theta} \right)^3 \cdot \int_0^{\theta/T} e^x x^4 (e^x - 1)^{-2} dx$ (θ is characteristic Debye temperature). Expression for polarization $\overline{P}_3 (R,r,T)$ was taken from Eq.(5a) neglecting the small contribution of thermal expansion for the sake of simplicity. $T_{cr}(R,r)$ is given by Eq.(6). Estimations show that in the vicinity of the size-driven phase transition point $T \approx T_{cr}(R,r)$ and for materials with β < 0, the efficiency tends to the maximal efficiency of Carnot circle: $\eta(R,r,T) \to 1 - T/T_{cr}(R,r)$.

Contour map of efficiency in coordinates temperature-radius and its radius dependence are shown in Figs. 4a,b. The efficiency is about several percents at room temperatures. Such low efficiency prevents direct application of ferroelectric nanowires as the heat power converter at normal conditions. Only at low temperatures the heat conversion into electric power by nanowires may be reasonable.

The current power frequency spectrum $\tilde{J}_{pr}^n(\omega)$ generated it response to the temperature variation $\delta \dot{T}(\omega)$ is shown in Figs. 4c for different wire radius $R$. It is clear that total pyroelectric charge $Q_{pr} = \tilde{J}_{pr}^n(\omega = 0)$ as well as the broadest spectrum correspond to the wire radius close to the critical one. The total charge $Q_{pr}$ decreases and tends to constant value with the wire radius increase (compare different curves in plot c).

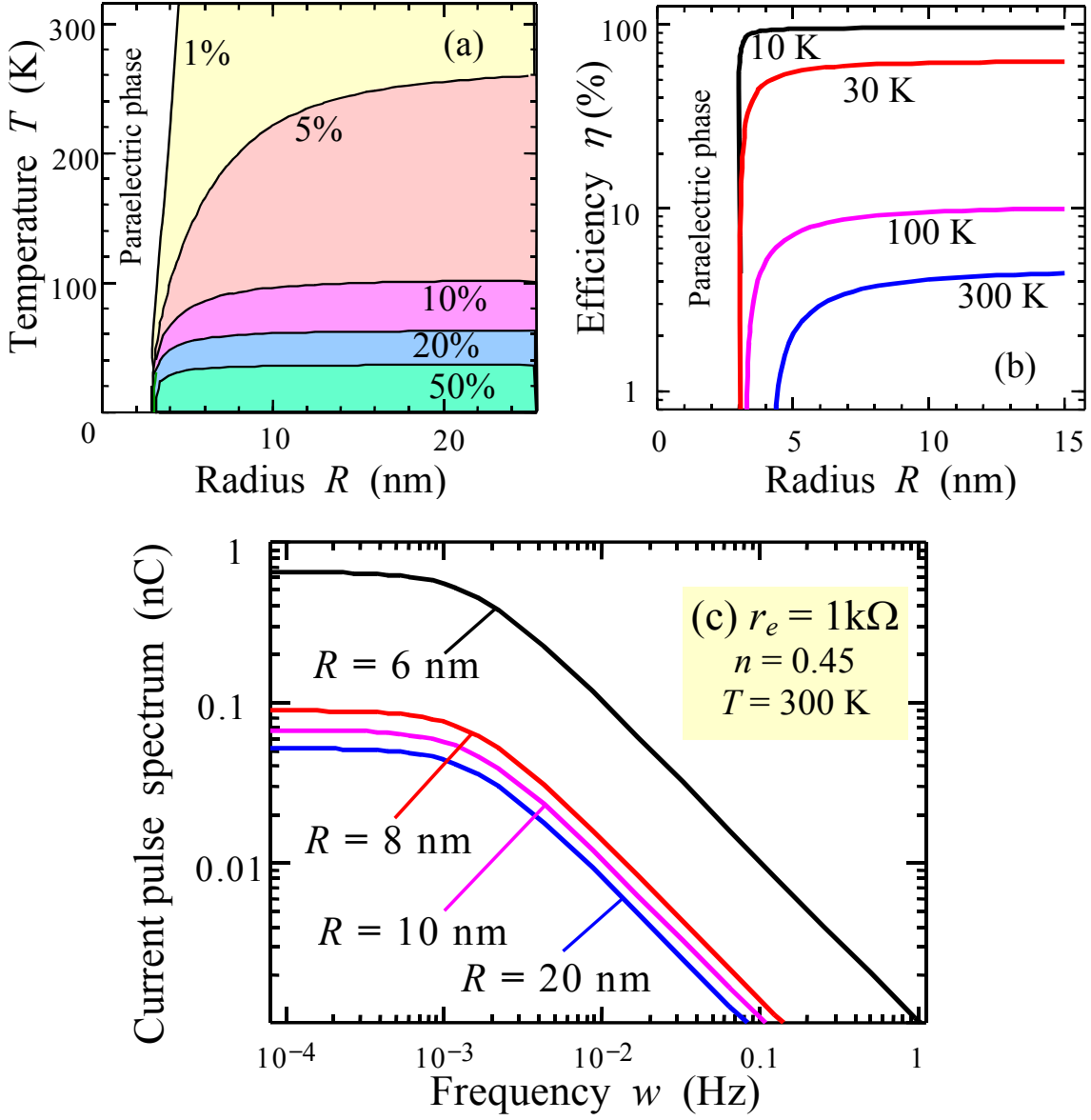

**FIG. 4.** (Color online) (a,b) Efficiency of the power converter vs. temperature and wire radius for parameters $\lambda = 1$ nm, $\mu^S = 1$ N/m, $\theta = 400$ K, $C_B = 3 \cdot 10^6$ J/(K·m³). (a) Contour map of η values in coordinates temperature-thickness. Different curves correspond to the fixed efficiency values 1, 5, 10, 20 and 50 %. (b) Efficiency vs. radius for different temperatures $T = 10, 30, 100, 300$ K (figures near the curves). (c) The current $\tilde{J}_{pr}^n(\omega)$ power spectrum at $T = 300$ K. Other parameters are the same as in Figs. 2 and 3a.

**Discussion**

Using concrete example of nanotubes and nanowires, we consider the influence of size effect on pyroelectric response of ferroelectric nanoparticles within phenomenological theory.

We obtained that pyroelectric coefficient increases with the wire radius decrease and diverges at critical radius $R_{cr}$ corresponding to the size-driven transition into paraelectric phase. Our analytical results predict that it is possible to tune pyroelectric coefficient value by varying the nanowire radius $R$ and ambient media (e.g. template material, gas or gel), since the ferroelectric-ambient interface determines the surface energy coefficient $\alpha_S$. The strong size effect on pyroelectric response should appear for arbitrary nanoparticle shape in the case when any of its sizes approach the critical ones.

We calculated that pyroelectric voltages $U_{pr} \sim 0.1$ V (in open-circuit mode) and direct current density $j_{pr} \sim 0.5$ nA/mm$^2$ (in short-circuit mode) can be generated by polarized Pb(Zr,Ti)O$_3$ ferroelectric nanowires array in response to a temperature variation with rate $dT/dt \sim 0.1$ K/s. The advantage of the proposed ferroelectric nanowire-based device is the absence of moving parts possible due to their pyroelectric response. The appropriate choice of the electrodes (e.g. silicon covered Au or Pt, or LSMO) and ferroelectric-semiconductor with definite electronic properties (e.g. donor-doped BaTiO$_3$, BiFeO$_3$, Pb(Zr,Ti)O$_3$, S$_2$P$_2$(S,Se)$_6$) make it possible to design Schottky barrier at the ferroelectric-semiconductor-metal interface. Rectification effect of the Schottky barrier allows the application of ferroelectric nanowires array fixed between the flat electrodes as the direct current generator.

Due to the size effect ferroelectric nanowires can successfully operate in the high-sensitive pyroelectric sensors, if the halfwidth $\Delta R$ of the nanowire radius distribution function is small enough: $\Delta R \ll \langle R \rangle$. The scattering in the radius $R$ unavoidably leads to noticeable diffuseness of the transition temperature $T_{cr}$ and strongly smears the pyroelectric coefficient size dependence.

The absence of moving parts makes capacitors filled with ferroelectric nanowires or nanotubes suitable for the harvesting of electric current and voltage from different heat sources.

The efficiency of pyroelectric nanoparticles used as the heat power converters into electric power is relatively low at room temperatures (about several %). However at temperatures close to the size-driven transition temperature $T_{cr}$ the efficiency tends to the maximal Carnot cycle efficiency, the latter can be increased at low temperatures (in particularly in the outer space).

**Acknowledgements**

Research sponsored by Ministry of Science and Education of Ukrainian and National Science Foundation (Materials World Network, DMR-0908718). EEA, ANM and GSS gratefully acknowledge financial support from National Academy of Science of Ukraine and Russian Academy of Science, joint Russian-Ukrainian grant NASU N 17-Ukr_a (RFBR N 08-02-90434). The research is supported in part (SVK) by the Division of Scientific User Facilities, DOE BES. Authors acknowledge multiple discussions with Profs. S.L. Bravina and N.V. Morozovskii.

**Appendix**

Using the Saint-Venant principle one could get the quantitatively correct physical picture (except the immediate vicinity of the faces $z = 0$ and $z = l$) and derived appropriate analytical expressions for the strain field in nanoparticles of tetragonal ferroelectric and cubic elastic symmetry

$$u_{11} + u_{22} \approx 2a_{11}^T(T - T_0) - 2\frac{1-\nu}{Y}\frac{\mu^S}{R} + u_c\left(1 - 2\frac{c_{12}}{c_{11}}\right)\frac{r^2}{R^2} + 2Q_{12}\left(1 - \frac{r^2}{R^2}\right)\overline{P_3^2} + \frac{q_{12}}{c_{11}}\left(P_3^2 - \left(1 - \frac{r^2}{R^2}\right)\overline{P_3^2}\right), \quad (A.1a)$$

$$u_{33} \approx a_{33}^T(T - T_0) + 2\frac{\nu}{Y}\frac{\mu^S}{R} + u_c\frac{r^2}{R^2} + \left(1 - \frac{r^2}{R^2}\right)Q_{11}\overline{P_3^2}, \qquad u_{23,12,13} = 0. \quad (A.1b)$$

Thermal expansion coefficients are denoted as $a_{ij}^T$ hereinafter. The tube outer radius is $R$, the inner radius is $r$ (see Fig. 1b); $u_c$ is the strain at the interface $\rho = r$ (if any). For the practically important case of the ferroelectric tube deposited on a rigid dielectric core, the tube and core lattices mismatch or the difference of their thermal expansion coefficients determines $u_c$ value (allowing for the possible strain relaxation for thick tubes). $Y$ is the Young module, $\nu$ is the Poisson coefficient. Parameters $Q_{12} = \dfrac{c_{11}q_{12} - c_{12}q_{11}}{(c_{11} - c_{12})(c_{11} + 2c_{12})}$ and $Q_{11} = \dfrac{(c_{11} + c_{12})q_{11} - 2q_{12}c_{12}}{(c_{11} - c_{12})(c_{11} + 2c_{12})}$ stand for the stress electrostriction coefficients. The first terms in Eqs.(A.1) originated from thermal expansion $\sim a_{ij}^T$, the second terms originated from intrinsic surface stress $\sim \mu^S$, the third terms are the strains induced by core $\sim u_c$, the last terms are the spontaneous strains created by inhomogeneous polarization. Hereinafter a dash over the letter stands for the averaging over the nanoparticle volume.

# References

[1] S.B. Lang, Pyroelectricity: From Ancient Curiosity to Modern Imaging Tool. Physics Today, **58**, № 8, 31-36 (2005).

[2] J.F. Scott. Data storage: Multiferroic memories. Nature Materials **6**, 256-257 (2007).

[3] Zhong Lin Wang and Jinhui Song, Piezoelectric Nanogenerators Based on Zinc Oxide Nanowire Arrays, Science, **312**, 242 (2006).

[4] Xudong Wang, Jinhui Song, Jin Liu, Zhong Lin Wang. Direct-Current Nanogenerator Driven by Ultrasonic Waves. Science, **316**, № 5821, 102 – 105 (2007).

[5] Yong Qin, Xudong Wang & Zhong Lin Wang, Microfibre–nanowire hybrid structure for energy scavenging, Nature Letters, **451**, 809, (2008).

[6] R.B. Olsen, D. Evans, Pyroelectric energy conversion: Hysteresis loss and temperature sensitivity of a ferroelectric material. J. Appl. Phys. **54**, 5941 (1983).

[7] R.B. Olsen, D.A. Bruno, and J. Merv Briscoe, Pyroelectric conversion cycles. J. Appl. Phys. **58**, 4709 (1985).

[8] A.S. Mischenko, Q. Zhang, J.F. Scott, R.W. Whatmore, N.D. Mathur. Giant Electrocaloric Effect in Thin-Film $PbZr_{0.95}Ti_{0.05}O_3$. Science **311** (5765), 1270 - 1271 (2006).

[9] A.S. Mischenko, Q. Zhang, R.W. Whatmore, J.F. Scott, N.D. Mathur, Giant electrocaloric effect in the thin film relaxor ferroelectric 0.9 $PbMg_{1/3}Nb_{2/3}O_3$–0.1 $PbTiO_3$ near room temperature. Appl. Phys. Lett. 89, 242912 (2006).

[10] Y. Luo, I. Szafraniak, N.D. Zakharov, V. Nagarajan, M. Steinhart, R.B. Wehrspohn, J.H. Wendorff, R. Ramesh, M. Alexe Nanoshell tubes of ferroelectric lead zirconate titanate and barium titanate. Appl. Phys. Lett. **83**, 440-442 (2003)

[11] Z.H. Zhou, X.S. Gao, J. Wang, K. Fujihara, and S. Ramakrishna, V. Nagarajan. Giant strain in $PbZr_{0.2}Ti_{0.8}O_3$ nanowires. Appl. Phys. Lett. **90**, 052902-1-3(2007)

[12] D. Yadlovker, S. Berger. Uniform orientation and size of ferroelectric domains. Phys Rev B **71**: 184112-1-6 (2005).

[13] D. Yadlovker, S. Berger. Ferroelectric single-crystall nano-rods grown within a nano-porous aluminium oxide matrix. Journal of Electroceramics, **22**, 281 (2009)

[14] A.N. Morozovska, E.A. Eliseev, and M.D. Glinchuk, Ferroelectricity enhancement in confined nanorods: Direct variational method. Phys. Rev. B **73:** 214106-1-13(2006)

[15] A.N. Morozovska, M.D. Glinchuk, E.A. Eliseev. Phase transitions induced by confinement of ferroic nanoparticles. Physical Review B. **76**, № 1, 014102 (2007).

[16] Yue Zheng, C.H. Woo, and Biao Wang. Surface tension and size effects in ferroelectric nanotubes. J. Phys.: Condens. Matter **20,** 135216 (2008).

[17] Yue Zheng, C.H. Woo, and Biao Wang. Pulse-loaded ferroelectric nanowires as an alternating current source. Nano Letters **8,** 3131 (2008).

[18] M.S. Majdoub, P. Sharma, T. Cagin, Dramatic enhancement in energy harvesting for a narrow range of dimensions in piezoelectric nanostructures. Phys Rev B **78**, 121407(R)-1-4 (2008).

[19] Z.-G. Ban and S. P. Alpay. Dependence of the pyroelectric response on internal stresses in ferroelectric thin films. Appl. Phys. Lett., **82** (20), 3499 (2003)

[20] M.D. Glinchuk, A.N. Morozovska, E.A. Eliseev. Ferroelectric thin films phase diagrams with self-polarized phase and electret state. J. Appl. Phys. **99**, 114102 (2006).

[21] Yue Zheng, C.H. Woo, Biao Wang, Controlling dielectric and pyroelectric properties of compositionally graded ferroelectric rods by applied pressure. Journal of Applied Physics, **101**, 116103 (2007)

[22] The thermal conductivity coefficient of the media $a^2=10^{-4}$ (metals), $10^{-6}$ (oxides), $10^{-8}$ (plastics) m$^2$/s, for $l=10^{-6}$ m one has $\tau_{FE} = (l/a)^2 = 10^{-6}$ sec for oxides. For $\tau=10^0$-$10^2$ sec and oxide materials temperature wavelength $\Lambda = 2\pi\sqrt{2a^2/\tau} = 10^{-3}$-$10^{-2}$ m.

[23] E. Fatuzzo, H. Kiess, R. Nitsche. Theoretical Efficiency of Pyroelectric Power Convertets. J. Appl. Phys. **37**, 510 (1966)